\documentclass{IEEEtran}
\usepackage{graphicx}
\usepackage{color}
\usepackage{amsmath}
\usepackage{esint}
\usepackage{mathtools}
\usepackage{amsfonts}
\usepackage{amssymb}
\usepackage{bigints}
\usepackage{xfrac}
\usepackage{relsize}
\usepackage{cite}

\begin{document}
\title{Efficient and Robust Polylinear Analysis\\of Noisy Time Series}
\author{Myrl~G.~Marmarelis}
\markboth{January 2016}{}

\maketitle

\begin{abstract}
A method is proposed to generate an optimal fit of a number of connected linear trend segments onto time-series data. To be able to efficiently handle many lines, the method employs a stochastic search procedure to determine optimal transition point locations. Traditional methods use exhaustive grid searches, which severely limit the scale of the problems for which they can be utilized. The proposed approach is tried against time series with severe noise to demonstrate its robustness, and then it is applied to real medical data as an illustrative example. The resulting identification of ``pivot'' points can find use in pattern recognition for system control problems.
\end{abstract}

\section{Introduction}
When observing time-series data, it is often of great utility to extract features that are of interest for a particular purpose. Such features include the locations of \textit{transition points}~\cite{Comparability,DNA,Ethiopia,Wagner}---points where there is a change in the perceived trend in the evolution of the time series. With the algorithm presented in this paper one is able to efficiently identify the most important piecewise linear trends present in a time series, at different resolutions. Previously proposed methods look for transition points through exhaustive search procedures~\cite{Lerman}, which tend to be time-consuming. In this paper, an approach termed \textit{polylinear analysis} is presented that makes use of a stochastic search algorithm to efficiently obtain high-density piecewise-linear least-squares fits while maintaining its robustness.

\subsection{State of the Art: Segmented Linear Regression}
In the medical community, the term \textit{segmented linear regression} encapsulates a plethora of different techniques. In general, these techniques are concerned with fitting onto a time series a continuous function composed of multiple linear trend segments. The imposition of continuity makes this problem significantly more difficult than similar problems without such a constraint.

Many studies are concerned with fitting a model of which both the locations and the number of so-called transition points are known~\cite{Ethiopia,Wagner}. Such studies would include those that aim to observe the effects of an intervention in a system by looking at the different trends in a variable during fixed intervals: pre-, during, and post-intervention.

It is more common for the number of transition points (a.k.a joinpoints) in a given data set to be unknown a priori. Many people have devised statistical methods to estimate how many should be placed~\cite{Kim,Comparability,DNA}. The prevailing way to determine the optimal locations of these transition points is through a grid search~\cite{Lerman}. It is simple to find an optimal fit once the locations of the transition points are given.

\subsection{Polylinear Analysis}
The novelty of the proposed method lies in its use of Simulated Annealing~\cite{SimulatedAnnealing} to find the optimal transition points, termed \textit{pivots}. The rationale is based on the observation that for most problems, a random search is much more efficient than the exhaustive grid search or manual search, yet still effective~\cite{Random}. The choice of the number of pivots is informed by certain characteristics of the data, and it can be tailored to the objectives of each study. A grid search is much too slow for large amounts of pivots; in the meanwhile, the proposed algorithm can efficiently handle any number of pivots. Henceforth the term ``pivot'' will be used to indicate a transition point or the first or last point of the polylinear fit. The term ``line'' will denote a linear trend segment between two adjacent pivots.

\section{Method}
The problem is formulated as follows: given a time series $\{v_t\}_{t=1}^n,$ we want to retrospectively fit $k$ connected line segments onto it so that the cost function, the standard square error, is minimized; the lines are defined through the connecting pivot points $\{(x_i,y_i)\}_{i=0}^k$. The model is fit over the entire data set, so we set $x_0 = 1$ and $x_k = n+1$. The reason for choosing $n+1$ instead of $n$ is due to the definition of the partial cost function in \eqref{eq:error} for the last line segment.

\subsection{Derivations}
Using the established notation we define the function representing each trend segment:
\begin{equation}
L_i(t) = \frac{y_{i-1}(x_i - t) + y_i(t - x_{i-1})}{x_i - x_{i-1}},\; t\in \left[x_{i-1},\,x_i\right]\!.
\end{equation}

\subsubsection{Pivot Height}
Ultimately the goal is to minimize the Sum of Squared Errors (SSE). The cost function is the sum of all the trend segments' SSEs:
\begin{subequations}
\begin{equation}
\varepsilon = \sum_{i=1}^{k} \varepsilon_i,
\end{equation}
\begin{equation}\label{eq:error}
\varepsilon_i = \sum_{t=x_{i-1}}^{x_i-1} (L_i(t) - v_t)^2.
\end{equation}
\end{subequations}
Now we can differentiate the error with respect to the height of a specific pivot. This way we can find the optimal height $y_i$ while keeping the locations and heights of its neighbors as well as its own location $x_i$ fixed. Notice only $\varepsilon_i$ and $\varepsilon_{i+1}$ depend on $y_i;$
\begin{equation} \label{eq:partials}
\frac{\partial \varepsilon}{\partial y_i} = \frac{\partial \varepsilon_i}{\partial y_i} + \frac{\partial \varepsilon_{i+1}}{\partial y_i}.
\end{equation}

Setting \eqref{eq:partials} to zero lets us derive an expression for $y_i$ in terms of its neighboring pivots and the data points captured by its trend segments. The simplified formula is presented below:
\begin{subequations}\label{eq:solution}
\begin{align}
\begin{split}
\varphi_i
&= \frac{1}{x_i-x_{i-1}}\sum_{t=x_{i-1}}^{x_i-1}\!v_t(t-x_{i-1})\\ &+ \frac{1}{x_{i+1}-x_i}\sum_{t=x_i}^{x_{i+1}-1}\!v_t(x_{i+1}-t)\\
&- \frac{y_{i-1}}{(x_i-x_{i-1})^2}\sum_{t=x_{i-1}}^{x_i-1}\!(x_i-t)(t-x_{i-1})\\ &- \frac{y_{i+1}}{(x_{i+1}-x_i)^2}\sum_{t=x_i}^{x_{i+1}-1}\!(x_{i+1}-t)(t-x_i),
\end{split}\\ \nonumber\\
\begin{split}
\psi_i
&= \frac{1}{(x_i-x_{i-1})^2}\sum_{t=x_{i-1}}^{x_i-1}\!(t-x_{i-1})^2\\ &+ \frac{1}{(x_{i+1}-x_i)^2}\sum_{t=x_i}^{x_{i+1}-1}\!(x_{i+1}-t)^2,
\end{split}\\ \nonumber\\
y_i &= \frac{\varphi_i}{\psi_i}.
\end{align}
\end{subequations}
The special cases for the first and last pivots are trivial to derive.

\subsection{Simulated Annealing}
While it is possible to derive a closed-form solution for an optimal $y_i$, such is not possible when looking for an optimal $x_i$. Instead, we need to search for the solution using an optimization algorithm. I chose to use simulated annealing for its relative speed and its ability to escape local minima~\cite{SimulatedAnnealing}. A solution is defined by the locations of the $k+1$ pivots. The neighborhood of a solution is then all the possible pivot sequences that have identical locations to those of the given solution except for exactly one. Simulated annealing works by walking through the neighborhoods while attempting to reduce the cost function. Each step has a ``temperature'' which determines both the potential size of the jump and the probability at which the system will move to a less optimal state than the current one. High temperatures are more volatile and allow the system to make big jumps in the solution space, even tolerating an increase in the cost function. By gradually reducing the temperature we expect that the system reaches a state that is close to the globally optimal solution.
\par The probability that a particular neighbor will be selected, at a given temperature $T$, is given by
\begin{equation}\label{eq:gamma}
G(x) = N\!\left(x|\;\mu = JT,\;\sigma = \sfrac{JT\!}{4}\right)
\end{equation}
where $N$ denotes the probability density function of the normal distribution and $x$ is the horizontal distance jumped by the displaced pivot point. The value of the standard deviation $\sigma$ was chosen empirically. $J$ is the maximum jump distance, which is usually proportional to the average number of data points per line. In other words, it is defined as $\gamma\frac{n}{k}$ where $\gamma$ is a constant usually less than 1.
\par The probability that a selected neighbor will be accepted as the new current solution comes from the physical inspiration of the algorithm and is thus given by the Boltzmann distribution:
\begin{equation}\label{eq:lambda}
P(\delta\varepsilon) = \mathrm{exp}\!\left(-\frac{\lambda\,\delta\varepsilon}{T}\right)
\end{equation}
where $\lambda$ denotes the strictness constant and $\delta\varepsilon$ is the change in error caused by the transition to the new state. $\lambda$ is adjusted empirically until a desired acceptance rate is reached.
\par The temperature starts at 1 and follows a geometric schedule; it is updated proportionally at each step:
\begin{equation}\label{eq:alpha}
T' = \alpha T
\end{equation}
where $\alpha$ is less than, but close to, 1. The temperature decreases monotonically, so the algorithm runs until $JT < 1$ because the $x$-values are integers and thus at this point the majority of jumps would be rounded to zero. Once this stage is reached the system is considered frozen and the best solution found so far is returned.

The initial state of the system comprises equally-spaced (in the horizontal axis) pivot points laid on the entire time series. Each step in the simulated annealing process marks a transition to a neighboring state (if it is accepted) and a reduction in the temperature. When a new state is chosen, the vertical locations of the displaced pivot point and those in its proximity need to be adjusted accordingly before evaluating the cost function. For this reason, once a neighboring solution has been selected, the $y$-values of all the pivots are updated iteratively using the closed-form solution \eqref{eq:solution} until the change in the error function becomes negligible.

\subsection{Choice of Parameters}
Since $\alpha$, in \eqref{eq:alpha}, controls the speed of the algorithm, small changes in its value could greatly influence the results. A faster run of the algorithm has increased reliance on the other parameters, $\lambda$ in \eqref{eq:lambda} and to a lesser extent $\gamma$ in \eqref{eq:gamma}, to produce quality fits. For noisy data it was found empirically that a near-optimal choice of strictness value $\lambda$ should cause the system to have an overall acceptance ratio of around 30\% to 35\%, though longer runs tend to have lower acceptance rates for the same $\lambda$. Most of the results shown below were generated with an $\alpha$ of 0.9997 (which corresponds to over 11000 generations) and a $\gamma$ of 0.8.

To get consistent results, one may need to run multiple simulations in parallel and then use the best result; otherwise, one may end up with the result of a simulation that started off in an ``unlucky'' fashion. A way to reduce the possibility of falling upon sub-par fits is to increase the length of the run: this causes it to spend more time in its initial volatile state, which is needed to avoid getting stuck in local optima.

\section{Results}

\begin{figure}
\includegraphics{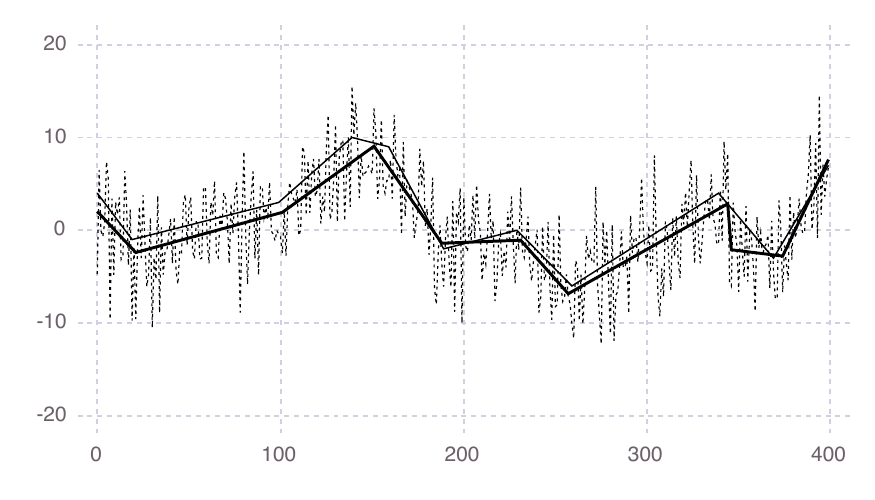}
\caption{Polylinear fit on a simulated time series with a signal-to-noise ratio (SNR) of 0 dB. Thick line is fit; thin line is original time series with (dotted) and without (solid) noise.}
\label{fig:simulate}
\end{figure}

\subsection{First, a Simulated Example}
To demonstrate the effectiveness of the aforementioned approach, a time series was constructed from ten trend segments plus noise. The signal-to-noise ratio (SNR) is defined as follows:
\begin{equation}
\mathrm{SNR_{dB}}=10\log_{10}\left(\frac{\sigma^2\mathrm{(signal)}}{\sigma^2\mathrm{(noise)}}\right)\!.
\end{equation}
Figure \ref{fig:simulate} shows the algorithm's performance on a time series with an SNR of 0 dB. The noise is white Gaussian noise. As can be seen, the pivot points are approximated by the algorithm. The exception lies on the shallow trend at around the time 140-150 that is insignificant compared to its neighbors in the noiseless version of the time series. The extra line that the model failed to put in the right place is then used to pick up some of the noise at around the time 350.

\begin{figure}
\includegraphics{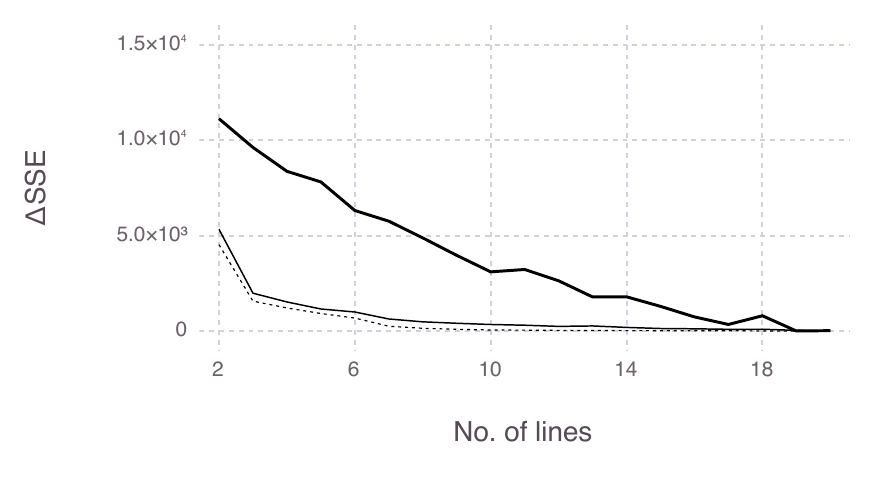}
\caption{Errors for polylinear fits of different resolutions (i.e. different numbers of lines). Ten should be optimal because simulation started with ten lines. Different lines correspond to simulations with different SNRs: thick is -10 dB; thin is 0 dB; dotted is 10 dB.}
\label{fig:trials}
\end{figure}

In Figure \ref{fig:trials} one can see the effects of changing the number of lines in the fit. To maintain a consistent acceptance rate, the strictness parameter $\lambda$ needs to increase linearly or super-linearly with respect to the number of lines; how much so depends on the nature of the data. For this example a binomial curve was derived with a constant plus a term with a fractional exponent in order to model this relationship. The -10 dB fit had an exponent of 1, the 0 dB an exponent of 1.5, and the 10 dB an exponent of 2.

\begin{figure}
\includegraphics{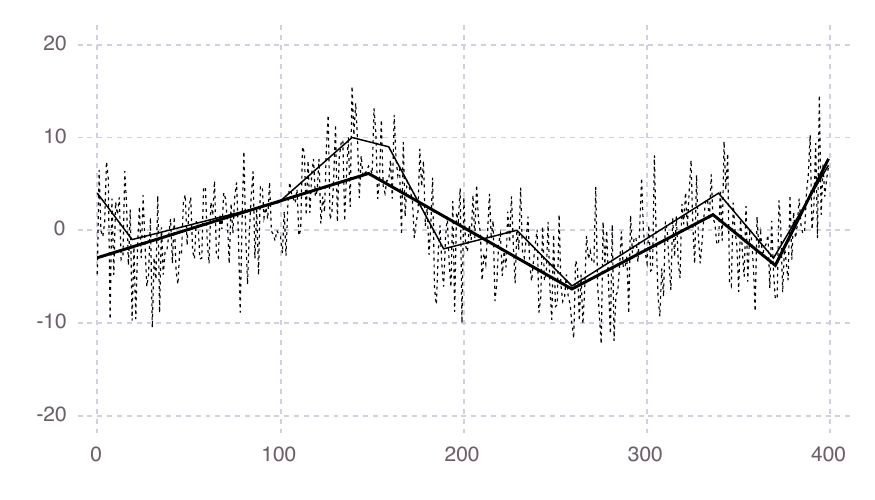}
\includegraphics{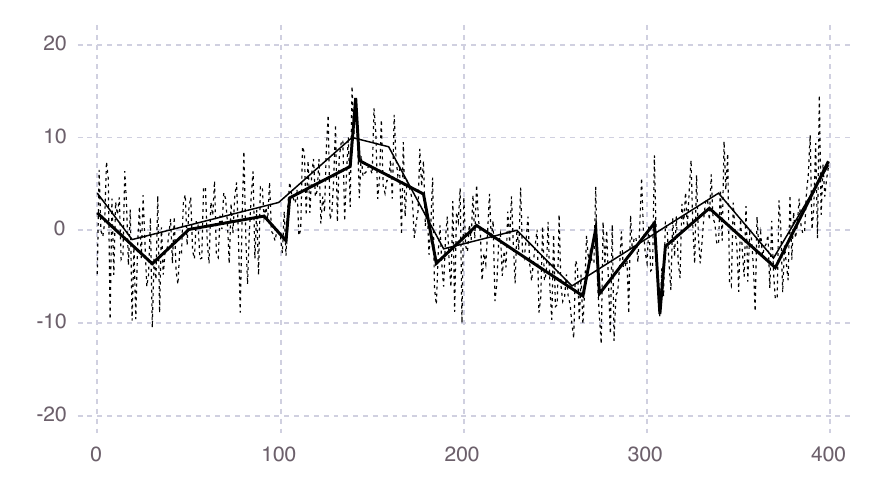}
\caption{Polylinear fits with half the number of lines (top) and double the number of lines (bottom) as the original time series that is being fitted upon. SNR is 0 dB; thick line is the fit and thin line is the simulated data, with (dotted) and without (solid) the noise.}
\label{fig:resolutions}
\end{figure}

\begin{figure}
\includegraphics{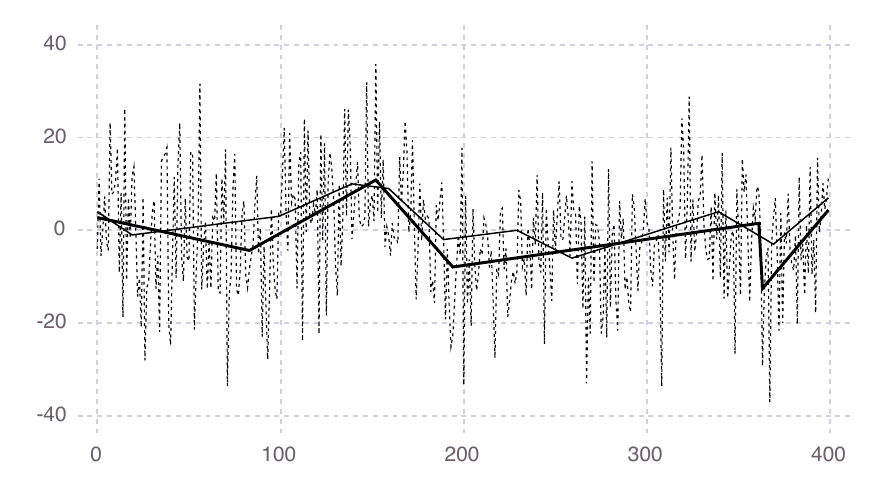}
\includegraphics{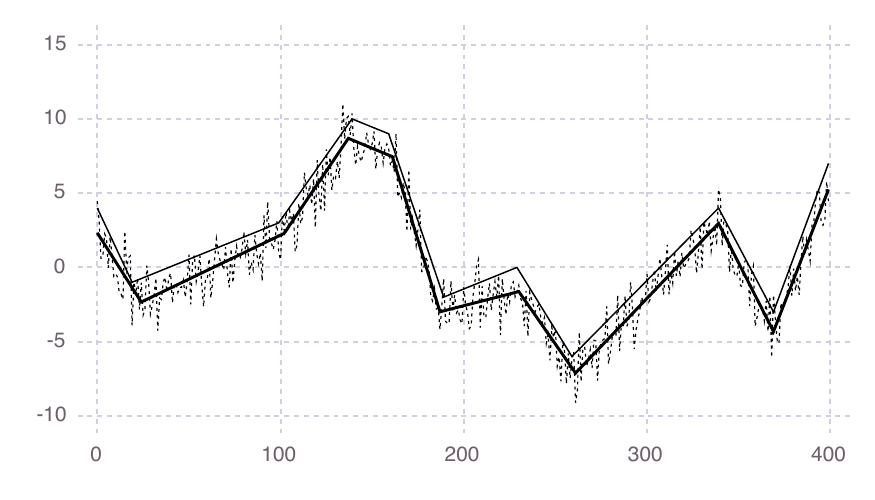}
\caption{Polylinear fits on simulations with different noise levels: SNRs of -10 dB (top) and 10 dB (bottom). The -10 dB fit has six lines instead of ten as it is favorable to under-specify the resolution under the presence of extreme noise. Thick line is the fit overlain on the simulated data with (dotted) and without (solid) the noise.}
\label{fig:noise}
\end{figure}

If the fit consists of less trend lines than are naturally present in the data, then it tends to ``merge'' the adjacent trends that are less distinguishable. Conversely, adding a higher number of lines than the underlying trends in the data causes the algorithm to overfit. These phenomena are evidenced in Figure \ref{fig:resolutions}. While fitting noise does reduce the overall error, it does not reduce it by much for fits with reasonable SNR. In fact, even before reaching ten lines, the impact on the error of adding more lines to the fit diminishes rapidly for the 0 and 10 dB data (as shown in Figure \ref{fig:trials}). One possible explanation is because some trend lines in the original simulation are not significantly different from their neighbors. Figure \ref{fig:noise} explores the performance of the algorithm under different noise levels. One can see that in the example with low noise the algorithm has managed to pick up every pivot point. Polylinear analysis is still a robust tool even under severe noise levels, though the resolution may need to be reduced from the theoretical optimum of the data in order to avoid fitting noise. In the noisy example in Figure \ref{fig:noise} one can still see that the lower-resolution fit matches the overall trends of the simulated data.

\subsection{An Illustrative Application to Real Data}
The proposed algorithm was applied to real time-series data beat-to-beat measurements of cerebral hemodynamics of three healthy humans. The data consists of two distinct, noninvasive measurements: of blood flow velocity at the left middle cerebral artery acquired via transcranial Doppler, and of arterial blood pressure measured via photoplethysmography at the index finger. The specifics of the data collection procedure and pre-processing to get de-meaned and resampled data (at 0.25 Hz) are given by \cite{Vasilis}. The data were provided by Dr. R. Zhang (see Acknowledgment). In previous studies these two variables are thought to interact through the cerebral auto-regulation process~\cite{Lagopoulos,Mitsis}, where the pressure is seen as the input and flow velocity as the output~\cite{Vasilis}. Henceforth the potential utility of polylinear analysis will be explored in terms of precise estimation of the latency between the two signals, which may attain physiological (and potentially clinical) significance.

\subsection{Choosing the Resolution of the Polylinear Fit}
\begin{figure}
\includegraphics{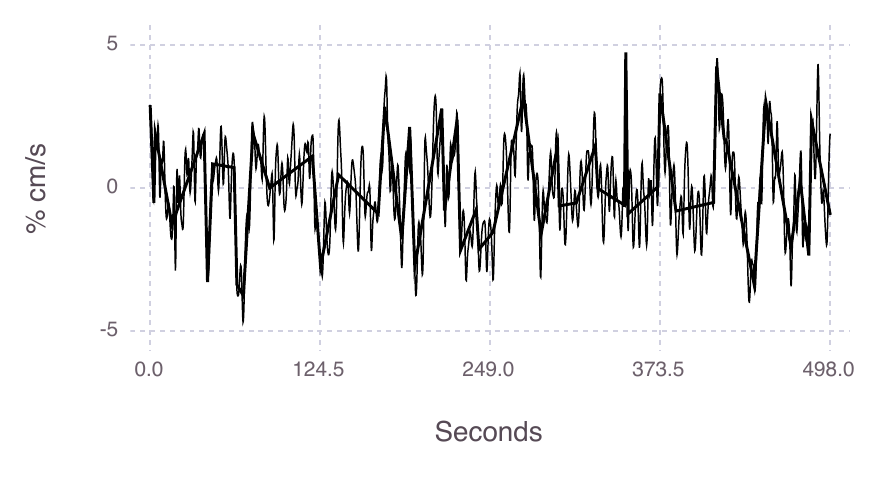}
\includegraphics{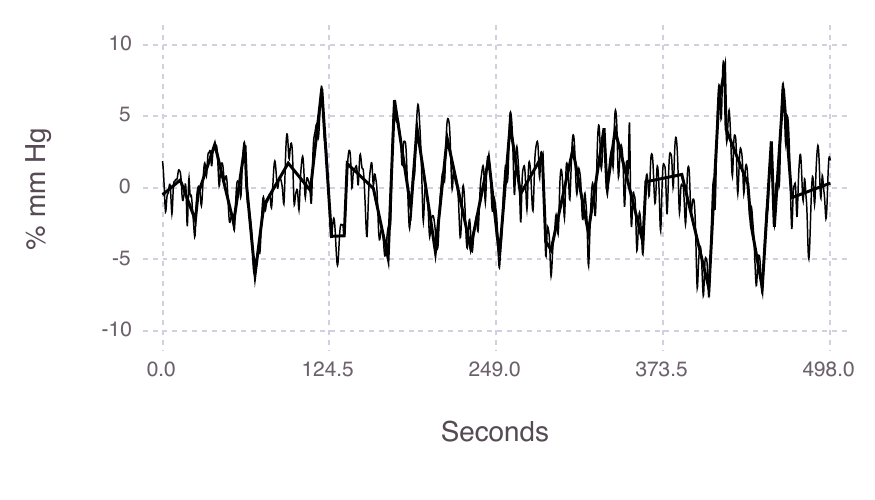}
\caption{Polylinear fits on a flow velocity time series (top) and the corresponding pressure time series (bottom). Thick line is fit and thin line is raw data, mean-normalized.}
\label{fig:fit}
\end{figure}

In order to amplify the features of the raw signal that are of interest to this study, the peak frequency from the data spectrum was found. The fit then would have two lines per one period of that sinusoid. For the present data this meant an initial line width of ten seconds, so the total number of lines in one of the time series is $k\nobreak=\nobreak\frac{\text{(\# of data points)}(0.25\sfrac{\text{sec}}{\text{data point}})}{10\text{sec}}.$ Figure \ref{fig:fit} shows two sample fits. With traditional methods that employ an exhaustive search mechanism to find pivots, it would be too time-consuming to generate the aforementioned fits.

\section{Discussion}

\begin{figure}
\includegraphics{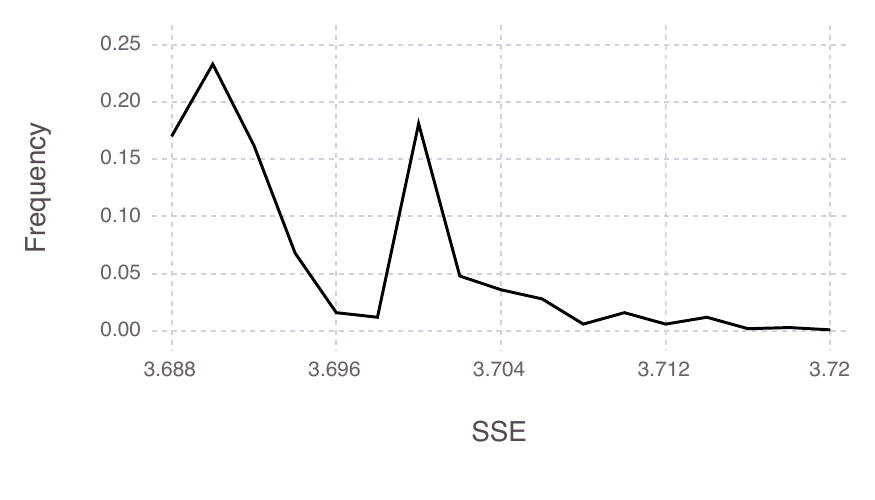}
\caption{Histogram of standard errors of 1000 fits on the 0 dB simulated data.}
\label{fig:probability}
\end{figure}

\subsection{Probability of Obtaining an Optimal Fit}
One thousand ten-line polylinear fits with identical parameters were generated on the simulated time series with the 0 dB SNR. Figure \ref{fig:probability} shows the approximate probability density function of the SSE values of these fits.

Out of those 1000 fits, the best solution generated had an error of 3.689. The probability of a particular fit's SSE landing in between 3.689 and 3.692 (roughly corresponding to the peak of the leftmost bump) is about 55.7\%; thus with six simulations in parallel, there is an overwhelmingly high chance of obtaining at least one fit within that range: 99.2\%, to be precise.

The histogram in Figure \ref{fig:probability} is bimodal probably due to a rather convincing local minimum that some runs of the algorithm fail to escape, with the leftmost bump corresponding to the near-global minimum.

\subsection{Polylinear vs. Fourier Analysis}
In biometrics, data is usually cyclical but not periodic. Polylinear analysis does not assume periodicity and as such it may be more appropriate than Fourier analysis in its application to biomedical problems.

\begin{figure}
\includegraphics{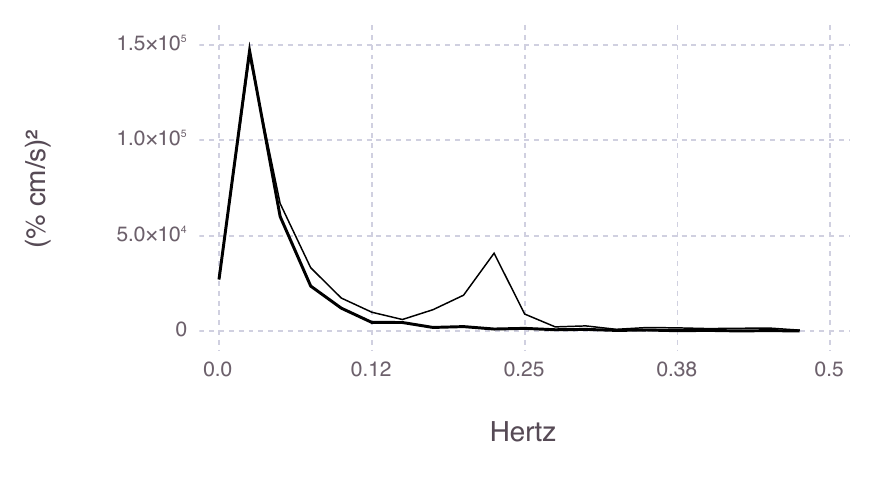}
\includegraphics{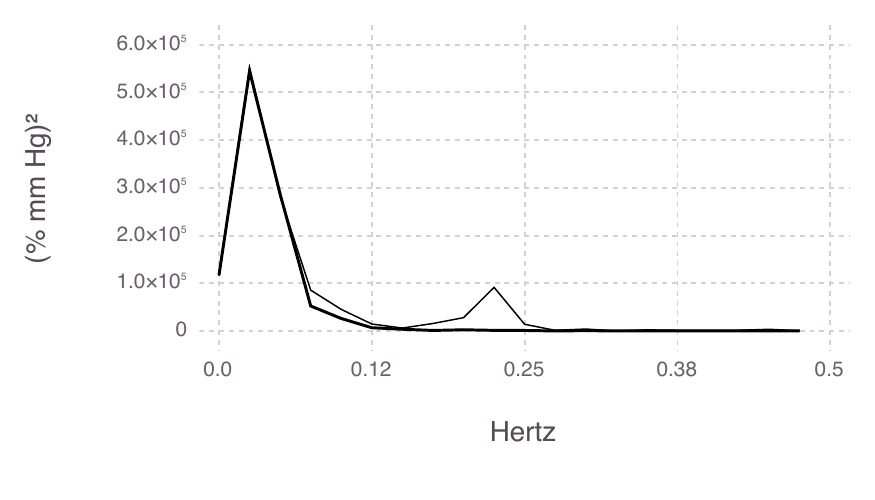}
\caption{Power Spectral Densities of flow velocity time series (top) and pressure time series (bottom). Thick line is of polylinear fit and thin line is of raw data.}
\label{fig:fft}
\end{figure}

The Power Spectral Density (PSD) is obtained by taking the Fourier transform of a function's autocorrelation. The autocorrelation of a time series $V$ is given by
\begin{equation}\label{eq:auto}
R_V(\tau) = \frac{1}{n\!-\!\tau}\sum_{t=1}^{n-\tau} V(t)V(t+\tau),
\end{equation}
and thus the PSD is
\begin{equation}
S_V(\omega) = \sum_{\tau = -\Delta\tau}^{\Delta\tau}\!\!\!R_V(\tau)\,\mathrm{e}^{-2\pi i \omega\tau},
\end{equation}
where $\Delta\tau$ is a user-defined maximum time lag. This parameter allows the resolution of the discrete Fourier tranform to be adjusted. We generally want $\Delta\tau$ to be short enough to avoid the noise effects that are more pronounced for large lags. In this case, the normalization in \eqref{eq:auto} does not have much impact either. Figure \ref{fig:fft} shows the PSDs of sample raw data and of their corresponding polylinear fits. The main peak in the lower frequencies is preserved but the higher frequency one is discarded by the model, acting as a sort of low-pass filter similar to Fourier-style filtering, but without biasing the low-frequency content. This effect is valuable to the present study since the higher frequencies correspond to other processes such as breathing and hormonal regulation~\cite{Panerai}, which contribute to latency estimation errors.

\subsection{Cross-Correlations}
The cross-correlation of two time series is defined as
\begin{equation}
(V_1 \star V_2)(\tau) = \sum_{t=1}^{n-\tau} V_1(t)V_2(t+\tau),
\end{equation}
where $V_1$ and $V_2$ are time series. The cross-correlation is traditionally used to estimate the latency between two signals by finding the time lag of the peak value.

\begin{figure}
\includegraphics{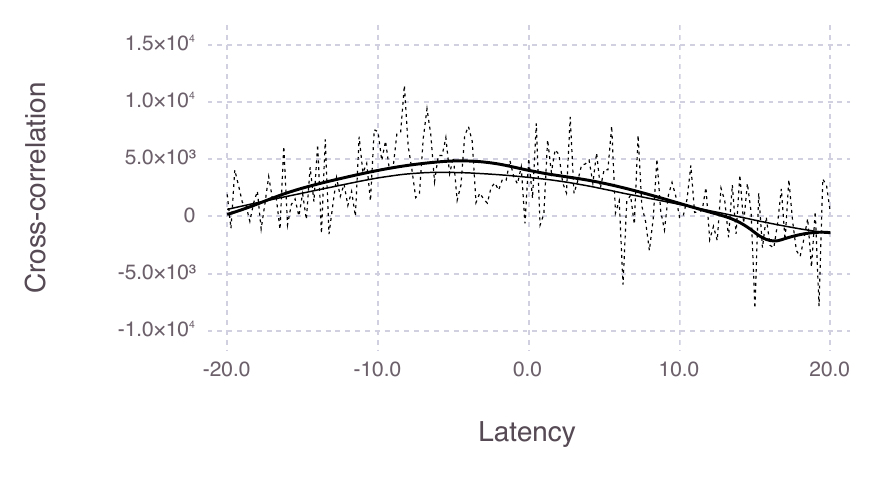}
\caption{Cross-correlations of simulated data that is time-shifted by 20 points, on both of which enough pseudo-white noise is added to set the SNR to -10 dB. Latency axis mimics the sampling rate of the real data: 4 points per time unit. Thick line is of fits; thin line is of heavily smoothed noisy signals; dashed line is of raw noisy signals.}
\label{fig:latency}
\end{figure}

\begin{figure}
\includegraphics{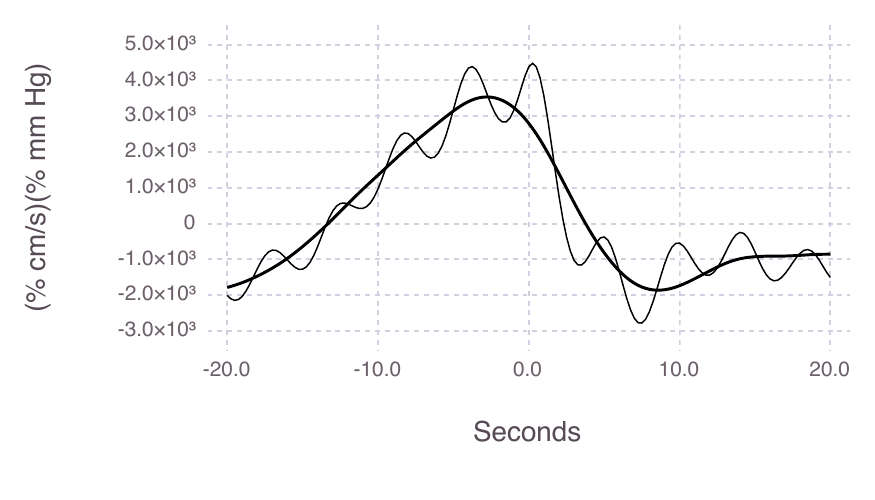}
\includegraphics{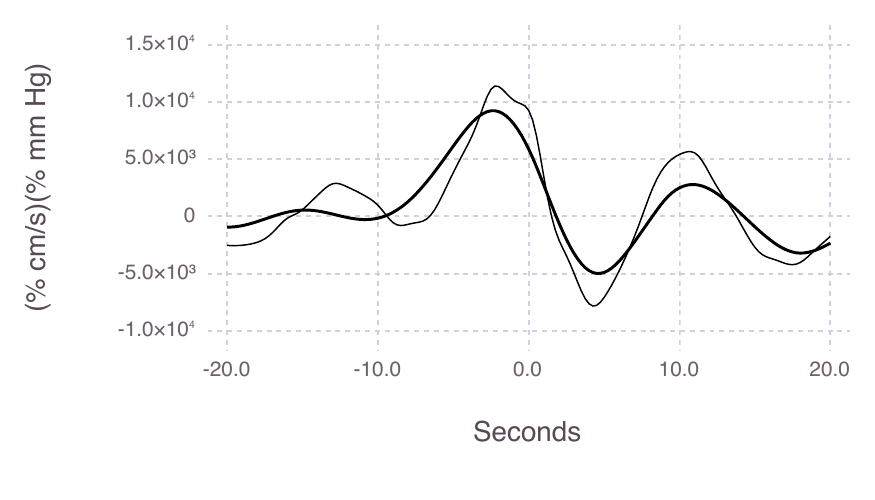}
\includegraphics{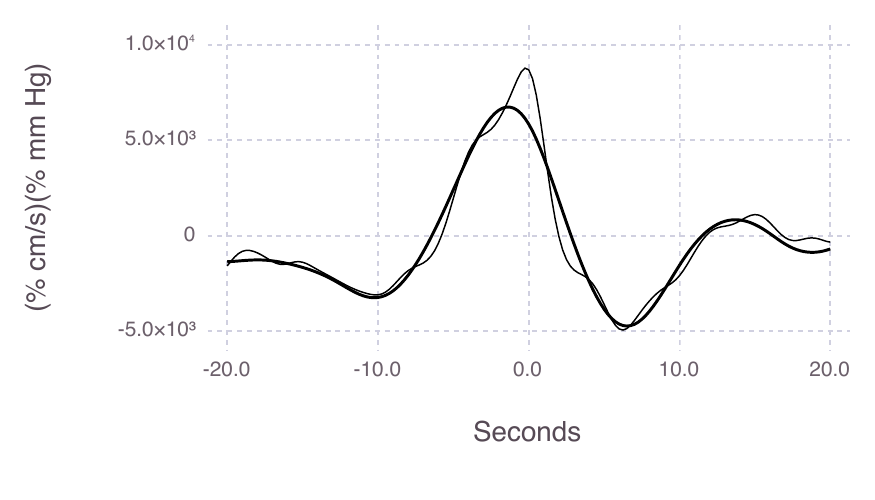}
\caption{Cross-correlations of the flow velocity and pressure measurements taken from the three subjects. Thick line is of fits and thin line is of raw signals.}
\label{fig:correlate}
\end{figure}

To demonstrate the efficacy of polylinear analysis in computing latency we first fabricate a scenario using the simulated data: the time series was cross-correlated with a replica shifted by 20 points (both embedded with independent -10 dB SNR noise). We want to observe the resilience of this method under extremely noisy conditions. Six lines (so with initial widths of $63.\bar{3}$) were fit on each of the time series and the result of this cross-correlation was compared with what one would obtain after applying a flat moving average of 59 points. The results are plotted in Figure \ref{fig:latency}.

The result obtained from the polylinear fits has a peak closer to the ground truth (just 1 point to the right) than that from the smoothed noisy data (4 points to the left). Further smoothing makes the curve so flat that it becomes difficult to distinguish the peak. Less smoothing shifts the peak more to the left; therefore in this instance polylinear analysis yields a better estimate of latency.

The cross-correlations of the polylinear fits can be used to estimate the latency between changes in blood pressure and blood flow velocity. Such is exemplified in Figure \ref{fig:correlate}, where the input-output (pressure-flow) cross-correlations are shown for three subjects. The average (standard deviation) values are -1.83 sec (0.76 sec) for polylinear analysis and -0.75 sec (1.32 sec) for raw data. The estimates obtained from cross-correlating the raw data have nearly twice the standard deviation as those from the polylinear fits. Hence the latencies can be obtained more clearly using the fits, since in the raw cross-correlations there are multiple peaks in different locations from those of the fits. The fits' peaks in Figure \ref{fig:correlate} closest to the zero point have horizontal displacements of -2.5 sec, -2 sec, and -1 sec, from top to bottom respectively. These findings agree with those of previous studies~\cite{Panerai}.

Polylinear analysis can prove useful for the estimation of latency and therefore assist in the diagnosis of various illnesses.

\subsection{Scalability}
Given a set of parameters for the simulated annealing, the speed of this algorithm depends only on the number of data points $n$: it scales as $\mathcal{O}(n)$ regardless of the number of lines~$k$. An exhaustive search, in contrast, scales as $\mathcal{O}\binom{n}{k-1}$. The fit does not need to be perfect and thus we can use a stochastic approach to quickly cover a large part of the search space~\cite{Random}.

\subsection{Comparison to Low-Pass Filtering}
Fourier analysis and its associated filtering methods can be used to achieve similar results; however, such an approach implicitly relies on the assumption that the low-frequency content remains unaltered, which is not guaranteed in Fourier-based low-pass filtering. On the other hand, polylinear analysis does not affect the low-frequency content appreciably.

\section{Conclusion}
In this paper an efficient and robust method is proposed for finding optimal least-squares fits of connected linear trend segments in time series. It was demonstrated that the proposed method maintains relatively accurate generation of results even in the presence of severe noise, and that these fits can be used reliably for determining the latency between two signals. The stochastic approach to searching for pivots allows it to greatly increase the number of fitted pivots with very little performance penalty.

\section*{Acknowledgment}
The author wishes to thank Dr. R. Zhang for providing the experimental data to test the proposed algorithm.

\bibliographystyle{ieeetr}
\bibliography{references}

\begin{thebibliography}{10}

\bibitem{Comparability}
H.-J. Kim, M.~P. Fay, B.~Yu, M.~J. Barrett, and E.~J. Feuer, ``Comparability of
  segmented line regression models,'' {\em Biometrics}, vol.~60,
  pp.~1005--1014, December 2004.

\bibitem{DNA}
S.~Nemes, T.~Z. Parris, A.~Danielsson, M.~Kannius-Janson, J.~M. Jonasson,
  G.~Steineck, and K.~Helou, ``Segmented regression, a versatile tool to
  analyze mrna levels in relation to dna copy number aberrations,'' {\em Genes,
  Chromosomes \& Cancer}, vol.~51, pp.~77--82, 2012.

\bibitem{Ethiopia}
T.~G. Gebrehiwot, M.~S. Sebastian, K.~Edin, and I.~Goicolea, ``The health
  extension program and its association with change in utilization of selected
  maternal health services in tigray region, ethiopia: A segmented linear
  regression analysis,'' {\em PLoS ONE}, 2015.

\bibitem{Wagner}
A.~K. Wagner, S.~B. Soumerai, F.~Zhang, and D.~Ross-Degnan, ``Segmented
  regession analysis of interrupted time series studies in medication use
  research,'' {\em Journal of Clinical Pharmacy and Therapeutics}, vol.~27,
  pp.~299--309, 2002.

\bibitem{Lerman}
P.~M. Lerman, ``Fitting segmented regression models by grid search,'' {\em
  Journal of the Royal Statistical Society, Series C}, vol.~29, no.~1,
  pp.~77--84, 1980.

\bibitem{Kim}
H.-J. Kim, M.~P. Fay, E.~J. Feuer, and D.~N. Midthune, ``Permutation tests for
  joinpoint regression with applications to cancer rates,'' {\em Statistics in
  Medicine}, vol.~19, pp.~335--351, 2000.

\bibitem{SimulatedAnnealing}
K.~A. Dowsland and J.~M. Thompson, ``Simulated annealing,'' in {\em Handbook of
  Natural Computing} (G.~Rozenberg {\em et~al.}, eds.), Springer, 2012.

\bibitem{Random}
J.~Bergstra and Y.~Bengio, ``Random search for hyper-parameter optimization,''
  {\em Journal of Machine Learning Research}, vol.~13, pp.~281--305, 2012.

\bibitem{Vasilis}
V.~Z. Marmarelis, D.~C. Shin, M.~E. Orme, and R.~Zhang, ``Time-varying dynamic
  modeling of cerebral hemodynamics,'' {\em IEEE Transactions on Biomedical
  Engineering}, vol.~61, pp.~1--11, 2014.

\bibitem{Lagopoulos}
J.~Lagopoulos, ``Cerebral autoregulation,'' {\em Acta Neuropsychiatrica},
  vol.~20, pp.~271--272, October 2008.

\bibitem{Mitsis}
G.~D. Mitsis, M.~J. Poulin, P.~A. Robbins, and V.~Z. Marmarelis, ``Nonlinear
  modeling of the dynamic effects of arterial pressure and co2 variations on
  cerebral blood flow in healthy humans,'' {\em IEEE Transactions on Biomedical
  Engineering}, vol.~51, pp.~1932--1943, 2004.

\bibitem{Panerai}
R.~B. Panerai, D.~M. Simpson, S.~T. Deverson, P.~Mahony, P.~Hayes, and D.~H.
  Evans, ``Multivariate dynamic analysis of cerebral blood flow regulation in
  humans,'' {\em IEEE Transactions on Biomedical Engineering}, vol.~47,
  pp.~419--423, March 2000.

\end{thebibliography}
\end{document}